\documentclass[aps,prl,twocolumn,superscriptaddress
,showpacs]{revtex4-1}
\usepackage{mathbbol,amsmath,amsfonts,amssymb,bm,color,dsfont}
\usepackage{graphicx,psfrag}

\def\dual{\ \stackrel{\Phi_\d}{\longrightarrow}\ }
\def\l{{\bm{l}}}
\def\p{{\bm{p}}}
\def\d{{{\sf d}}}
\def\r{{\bm{r}}}

\def\i{{\bm{e_1}}}
\def\j{{\bm{e_2}}}

\def\Z{{\mathbb{Z}}}

\def\sdual{\ \stackrel{\Phi_{\sf sd}}{\longrightarrow}\ }

\providecommand{\ignore}[1]{}

\makeatletter
\newcommand*{\balancecolsandclearpage}{%
  \close@column@grid
  \clearpage
  \twocolumngrid
}
\makeatother

\begin{document}

\title{Holographic Symmetries and Generalized Order Parameters for
Topological Matter}
\author{Emilio Cobanera}
\affiliation{Instituut-Lorentz, Universiteit Leiden, P.O. Box 9506, 2300
RA Leiden,  The Netherlands}
\email[Electronic address: ]{ecobaner@indiana.edu}
\author{Gerardo Ortiz} 
\affiliation{Department of Physics, Indiana University, Bloomington,
IN 47405, USA}
\author{Zohar Nussinov}
\affiliation{Department of Physics, Washington University, St.
Louis, MO 63160, USA}

\date{\today}

\begin{abstract}
We introduce a universally applicable method, based on the
bond-algebraic theory of  dualities, to search for  generalized order
parameters in disparate systems including non-Landau  systems with
topological order. 
A key notion that we advance is that of  {\em holographic
symmetry}. It reflects situations wherein global symmetries become,
under a duality mapping, symmetries that act solely on the system's boundary.  
Holographic symmetries are  naturally related to edge modes
and localization. The utility of our approach  is illustrated by
systematically deriving generalized order parameters for pure and
matter-coupled Abelian gauge theories, and for some models of 
topological matter.  
\end{abstract}
\maketitle

{\it Introduction.}--- Landau's concept of an order parameter (OP) and 
spontaneous symmetry breaking are central in physics
\cite{book}. In systems with long-range Landau orders, two-point
correlation functions of an OP  field ${\cal O}(\r)$, 
in their large distance limit, tend to a finite
(i.e., non-zero) value,
$\lim_{|\r-\r'| \rightarrow \infty}\lim_{N^d \rightarrow \infty} 
\langle {\cal O}(\r){\cal O}^{\dagger}(\r')\rangle \neq 0$,
where $N$ is the linear size of the $d$-dimensional system, and 
${\cal O}(\r)$ is {\it local} in
the (spatial) variable $\r$. It is in Landau's spirit to use the OP as a macroscopic variable 
characterizing the ordered phase {\it and} as an indicator of a possible
phase transition (classical or quantum) to a disordered state where the 
OP becomes zero.

There is much experience, including systematic methods
\cite{batista,furukawa},  for deriving Landau OPs and  their effective field
theories \cite{book}. Landau's ideas of a
({\it local}) OP 
cannot be extended to topological states of matter  as, by 
definition \cite{wen,TQO}, these lie beyond Landau's  paradigm. 
However,
the notion of  long-range
order, or the design of a {\it witness correlator} 
(i.e., a correlator discerning the existence of various phases and related transitions), 
can be extended to topological phases - phases that 
can only be meaningfully examined by {\it non-local probes} \cite{TQO}. 
Topological orders appear in
gauge theories, quantum Hall and spin  liquid states (when defined as deconfined phases of 
emergent gauge theories  \cite{sondhi}), 
including well-studied exactly solvable  models \cite{kitaev1,kitaev2}. 

In this paper  we demonstrate that generalized  {\it non-local} OPs may diagnose  
topological phases of matter. 
Most importantly, we
outline a method based on bond-algebraic duality mappings
to search systematically for
generalized OPs.
Dualities have the striking capability of mapping Landau 
to topological orders and viceversa for essentially two reasons.
First,  dualities in general represent  non-local
transformations of elementary degrees of freedom \cite{wegner}  and may even perform
transmutation of  statistics \cite{con}. Second, 
bond-algebra techniques \cite{con,nadprl,bond} allow for the
generation of dualities in finite and infinite size systems. As we will show, in 
systems with a boundary, dualities realize a form of  {\it
holography} \cite{zon} capable of transforming a {\it global} symmetry that may
drive spontaneous symmetry breakdown into a {\it boundary} symmetry. 
We term these distinguished boundary symmetries  {\it holographic}. 
They are under suitable further conditions 
connected to {\it edge (boundary) states}.
To illustrate the method, we derive explicitly a (non-local) witness
correlator  and a generalized OP, suited to diagnose the transition 
between deconfined and confined 
phases of matter-coupled 
gauge theories, undetectable by 
standard OPs or Wilson loops.
Other examples are reported in Ref \cite{SI}.


{\it The search for generalized order parameters.}--- 
A natural mathematical {\it language} to describe a physical 
system is that for which the system's degrees of freedom couple {\it locally}. 
This simple observation is key to understanding 
that topological order is a property of a state(s) relative to the 
algebra of observables (defining the language)
used to probe the system experimentally \cite{TQO}. 
In the language in which the system is topologically ordered, it is also robust
(at zero temperature \cite{thermal}) against 
perturbations local in that language. 
Spectral properties 
are invariant under unitary
transformations  of the local Hamiltonian $H$ governing the system,
\(H\mapsto UHU^\dagger\). If  \(UHU^\dagger\) corresponds to a sensible
local theory then the unitary transformation \(U\) establishes a {\it duality}
\cite{con}. A duality may map a system that displays topological order to one
that does not \cite{TQO}. Dualities for several of Kitaev's models 
\cite{kitaev1,kitaev2,kitaev_wire} epitomize this idea \cite{TQO,bond,thermal}.

Since dualities are unitary transformations (or, more generally, partial isometries)
\cite{con} they cannot in general change a phase diagram, only its interpretation. 
This leads to a central point of our work: {\it A duality mapping a Landau 
to a topologically ordered system must map the Landau OP to a
generalized OP characterizing the topological order.} Our
method for searching for generalized OPs, combines this observation with the 
advantages of the bond-algebraic theory of dualities \cite{con}. In this framework,
dualities in arbitrary size (finite or infinite) systems can be
systematically searched for as alternative local representations of
{\it bond algebras} of interactions associated  to a Hamiltonian \(H\).
Hence it is possible for {\it any} system possessing topological order  
to systematically search for a duality mapping it to a Landau
order. When a dual Landau theory is found, the dual system's OP can
be mapped back to obtain a generalized OP for the topologically
ordered system. 
In what follows and in Ref. \cite{SI}, we study various quantum gauge and 
topologically ordered theories, and their duals, to illustrate our ideas.


{\it Holographic symmetries and edge states: the gauged Kitaev wire.}--- 
We next illustrate  the concept of holographic symmetry and its relation to
generalized OPs and edge modes.  Consider  the  Kitaev wire
Hamiltonian \cite{kitaev_wire} with open boundary conditions, here generalized to 
include a \(\Z_2\) gauge field (termed the {\it gauged Kitaev wire}),
\begin{equation}
\label{HH2}
H_{\sf GK}=-ih \! \sum_{m=1}^Nb_ma_m- \! \sum_{m=1}^{N-1}[
iJb_m\sigma^z_{(m;1)}a_{m+1}+\kappa \sigma^x_{(m;1)}],
\end{equation} 
where \(a_m=a_m^\dagger\), \(b_m=b_m^\dagger\) denote two Majorana
fermions (\(\{a_m,a_n\}=2\delta_{m n}=\{b_m,b_n\}\), \(\{a_m,b_n\}=0\))
placed on each site of an open chain with \(N\) sites. The Pauli
matrices \(\sigma^\alpha_{(m;1)}, \alpha=x,z\), placed on the links
\((m;1)\) connecting sites  \(m$ and $m+1\) represent a \(\Z_2\) gauge
field. For the gauged Kitaev wire, fermionic parity is obtained
as the product of the local (gauge) \(\Z_2\) symmetries \(ib_1a_1\sigma^x_{(1;1)}\),
\(\sigma^x_{(N-1;1)}ib_Na_N\), and
\(\sigma^x_{(m;1)}ib_{m+1}a_{m+1}\sigma^x_{(m+1;1)}\)
(\(m=1,\cdots,N-2\)). Just like the standard Kitaev wire, \(H_{\sf
GK}[h=0]\) has two free edge modes  \(a_1\) and \(b_N\).

The gauged Kitaev wire holds two important dualities. It is dual
to the one-dimensional \(\Z_2\) Higgs model \cite{fradkin_shenker}
\begin{equation}\label{hhh}
H_{\sf H}=-h\sum_{i=1}^N\sigma^x_i-\sum_{i=1}^{N-1}[
J\sigma^z_i\sigma^z_{(i;1)}\sigma^z_{i+1}+\kappa \sigma^x_{(i;1)}],
\end{equation}
with Pauli matrices $\sigma_i^\alpha$ placed on sites $i$.
Moreover, the gauge-reducing \cite{con} duality mapping $\Phi_{\sf d}$
\begin{eqnarray} 
\hspace*{-0.5cm}\label{gz}
ib_m a_m  &\dual& \sigma^{z}_{m} \sigma^{z}_{m+1}, \quad m=1,\cdots,N,  \\  
ib_m \sigma^{z}_{(m;1)} a_{m+1} &\dual& \sigma^{x}_{m+1},  \quad m=
1,\cdots,N-1,  \nonumber \\ 
\sigma^{x}_{(m;1)} &\dual& \sigma^z_{m+1}, \quad m= 1,\cdots,N-1, \nonumber
\end{eqnarray} 
transforms $H_{\sf GK}$ into a spin-1/2 system 
\begin{equation}
H^D_{\sf
GK}=-h\sum_{m=1}^N\sigma^z_{m}\sigma^z_{m+1}-\sum_{m=2}^{N}[J\sigma^x_m
+\kappa \sigma^z_m].
\label{hdh}
\end{equation}
defined on $(N+1)$ sites. The fermionic parity $P$ maps to a 
holographic symmetry under this duality, since $
P=\prod_{m=1}^N ib_ma_m \!\! \dual \!\! \sigma^z_1\sigma^z_{N+1}
$, i.e., the product of two (commuting) boundary symmetries. 
Holography is a relational phenomenon (see \cite{SI}). 
A duality that uncovers a
holographic symmetry links a global (higher-dimensional) symmetry of
a system to a  boundary (lower-dimensional) symmetry of its dual.
Boundary symmetries need not  in general be duals of global symmetries.
 
What is the physical consequence of having an holographic symmetry? 
Consider the not uncommon situation in which the holographic symmetry is  
supplemented by an additional (non-commuting) boundary symmetry
{\it in some region of the phase diagram}. By definition, holographic 
symmetries are boundary symmetries which are
dual to global symmetries. Thus, global symmetries linking degenerate states (and properties in
the broken symmetry phase) in the dual system have imprints in their
holographic counterparts. 
Then, the many-body level degeneracy of the ground state 
may be ascribed  to boundary effects.  
If the couplings are now changed, the ground state degeneracy may get removed, 
together with some boundary symmetries. 
However, so long as the system remains in a topological phase dual
to the (broken symmetry) ordered phase, the low energy state splitting 
will be {\it exponentially small} in the system size, so that  in the thermodynamic limit ground state degeneracy is restored. 

The language providing the most local operator description of the ground-state manifold 
is the one realizing the edge modes, which are expected to be exponentially 
localized to the boundary. Thus, as long as the thermodynamic-limit degeneracy remains, 
a suitable local probe will detect localization on the boundary for those states. Conversely,
non-commuting edge mode operators in a gapped phase reflect the existence of low-energy
many-body states with energy splittings vanishing 
exponentially  in the system size. 
Many-body ({\it zero-energy}) edge states are thus simply a natural consequence 
of a degenerate ground state manifold in a gapped system.  
They are witnesses of an ordered (degenerate) 
phase described in a most local language. Note that 
boundary operators that commute with the Hamiltonian at special values of the
coupling(s)  
are a necessary but not sufficient condition to realize {\it exact} (zero-energy) edge modes. 

The duality \(H_{\sf H}\rightarrow H_{\sf GK}\)
maps a global symmetry of  \(H_{\sf H}[\kappa,h=0]\) to a boundary symmetry
of \(H_{\sf GK}[\kappa,h=0]\), i.e., \(\sigma^x_1\cdots\sigma^x_{N-1}\sigma^y_N\rightarrow b_N\), and
one boundary symmetry to another, 
\(\sigma^z_1\rightarrow a_1\).
If we now turn on \(h<J\), keeping 
\(\kappa=0\), the edge mode operators \(a_1,b_N\), evolve respectively into
\(\Gamma_1=\sum_{m=1}^N(-h/J)^{m-1}a_m(\prod_{s=1}^{m-1}\sigma^z_{(s;1)})\) 
and \(\Gamma_2=\sum_{m=1}^{N}(h/J)^{N-m} b_m(\prod_{s=m}^{N-1}\sigma^z_{(s;1)})\).
These modes are exponentially localized as long as the system is in 
the ordered gapped phase within a gauge sector  \cite{commutators}.  
The Majorana language affords a {\it local} boundary 
description of these (partly non-local in the Higgs language) zero-energy modes. 
For \(h>J\), and/or \(\kappa>0\), 
the ground state is unique, even in the thermodynamic limit, as we learn from the phase
diagram of the one-dimensional Higgs model \cite{fradkin_shenker}. 
Hence the zero-energy modes disappear together with the ground-state
degeneracy. For $\kappa > 0$, 
they disappear despite the fact that  fermionic 
parity remains an exact symmetry and cannot be spontaneously broken \cite{elitzur}. 
Consider now  
\(H_{\sf GK}^D\) of Eq. \eqref{hdh}. At \(h=0\), it has
zero-energy edge mode operators \(\sigma^z_1,\sigma^x_1,\sigma^z_{N+1},\sigma^x_{N+1}\). 
For $h>0$, and \(\kappa=0\), two of these remain unchanged, and the other
two evolve into \(\Sigma_1=\sigma^x_1+\sum_{m=1}^{N-1}(h/J)^m\sigma^y_1
(\prod_{s=2}^{m}\sigma^x_s)\sigma^y_{m+1}\) and \(\Sigma_2=
\sigma^x_{N+1}+\sum_{m=2}^{N}(h/J)^m\sigma^y_m
(\prod_{s=m+1}^N\sigma^x_s)\sigma^y_{N+1}\).
These  behave just as their Majorana relatives,  
yet they are recognized as non-local. 
The Majorana language distinguishes itself as the {\it most local one} for zero modes.   

To obtain a generalized OP for the gauged Kitaev wire, notice
that \(H^D_{\sf GK}[\kappa=0]\)
reduces to the transverse-field Ising ({\sf TI}) chain. 
Hence it exhibits a second-order phase
transition at \(J=h, \kappa=0\). For \(H^D_{\sf GK}[\kappa=0]\),
this transition is witnessed by the Landau OP correlator
\(\lim_{|i-j|\rightarrow \infty}\langle {\sf
TI}|\sigma^z_i\sigma^z_j|{\sf TI}\rangle\). (From now on 
\(|{\sf label}\rangle\) represents the ground state of \(H_{\sf label}\)).  
Our duality maps this correlator back to a generalized OP for the gauged Kitaev wire, 
the string correlator 
\(\lim_{|i-j|\rightarrow \infty}\langle {\sf
GK}|ib_ia_{i}ib_{i+1}a_{i+1}\cdots ib_ja_j|{\sf GK}\rangle\).

{\it Generalized OPs in higher-dimensional theories---} We next show 
how to systematically {\it derive} generalized OPs in higher space dimensions. 
Our main goal is to  
illustrate the methodology in the challenging case of  the Abelian (\(U(1)\)) 
matter-coupled gauge (Higgs) theory. 
Previous works \cite{bricmont,fm} {\it conjectured} generalized OPs for  
matter-coupled gauge theories and were numerically implemented, for instance, in 
Ref. \cite{sondhi}. Unfortunately, a systematic mathematical {\it derivation} 
was missing and this is what our work is about. 
Our (non-local) witness correlator for the Higgs model turns out to be  
the one conjectured in Ref. \cite{bricmont}.   In Ref. \cite{SI},  
we study 
several other examples (displaying also holographic symmetries), 
including Ising and \(\Z_p\) gauge and Higgs theories,
the \(\Z_p\) extended toric code \cite{vidal} as an interesting example of topological 
order, and the XY model on the frustrated Kagome lattice.
Non-Abelian extensions of our ideas based on Ref. \cite{nadprl}
are currently under investigation. 

\begin{figure}[h]
\includegraphics[angle=0,width=.58\columnwidth]{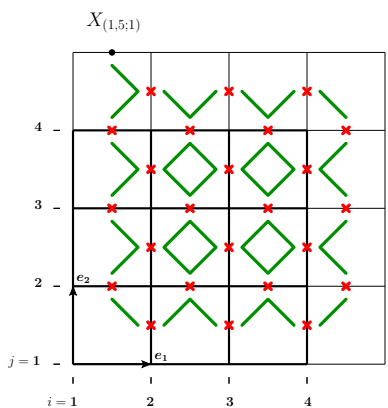}
\caption{The \(\mathds{Z}\) gauge theory exactly dual to the quantum XY
model must satisfy special boundary conditions and possesses a boundary
symmetry. The lattice corresponding to the  XY model is shown in thick
lines, for \(N=4\).}
\label{dual_ising_fig}
\end{figure}

To derive the generalized OP for the Abelian Higgs theory, 
our starting point is the XY model defined in terms
of continuous \(U(1)\) degrees of freedom 
\(s_{\r}\equiv e^{-i\theta_{\r}},\ \theta_\r\in[0,2\pi),\)
placed at sites $\r = i\i+j\j= (i,j)$ of a square lattice. 
The model's Hamiltonian reads 
\begin{eqnarray}
H_{\sf XY}&=&h\sum_{i,j=1}^N L_{(i,j)}^2  \label{hhxy}\\
&+&\frac{J}{2}\Big[
\sum_{i=1}^{N-1}\sum_{j=1}^{N}S_{(i,j;1)} 
+\sum_{i=1}^{N}\sum_{j=1}^{N-1}S_{(i,j;2)}+h.c.\Big] , \nonumber
\end{eqnarray}
with \(L_{\r}\equiv-i\partial/\partial\theta_{\r}\), and 
\(S_{(\r;\mu)}\equiv s_\r^{\;} s_{\r+\bm{e_\mu}}^\dagger\).  
The XY model is dual
to 
a \(\Z\)  (solid-on-solid like) gauge
theory 
also defined on a square lattice,  but
with degrees of freedom $X$ and  $R$ associated to links
($\r; \mu=1,2)$. (In matter-coupled gauge theories we will 
also have operators acting on sites \(\r\).) 
These operators  satisfy \(X |m
\rangle=m  |m \rangle\),  \(R|m\rangle= |m-1\rangle\), 
\(R^\dagger|m\rangle= |m+1\rangle\), with $m \in \mathds{Z}$, 
and commute on different  links (and/or sites).   
Then, the {\it exact} dual of \(H_{\sf XY}\) for {\it open
boundary conditions} reads
\begin{eqnarray}
\label{hzg}
H_{\sf ZG}&=& h\sum_{i,j=1}^N b_{(i,j)}^2 \\
&+&\frac{J}{2}\Big[\sum_{i=2}^N\sum_{j=1}^N R_{(i,j;2)} 
+\sum_{i=1}^{N}\sum_{j=2}^N R_{(i,j;1)} + h.c.\Big].\nonumber
\end{eqnarray}
We will call {\it system
indices} the link indices \((i,j;\mu=1,2)\)  labeling \(R\)
operators that explicitly appear in \(H_{\sf ZG}\), and  
{\it extra indices} the remaining link indices. In
the bulk, the plaquette operator \(b_{(i,j)}\) reads 
\begin{equation}\label{plaquette1}
b_{(i,j)}\equiv X_{(i,j;1)}+X_{(i+1,j;2)}-X_{(i,j+1;1)}-X_{(i,j;2)}.
\end{equation}
On the lattice boundary, the plaquette operators are set by  two rules: 
(i) \(b_{(1,N)}=X_{(1,N;1)}-X_{(2,N;2)}-X_{(1,N+1;1)}.\)
Thus, \(b_{(1,N)}\) involves one degree of freedom \(X_{(1,N+1;1)}\)
labelled by an extra  link index. (ii) The remaining boundary plaquettes
are determined by  Eq. \eqref{plaquette1} {\it provided operators
labelled by extra link indices are omitted}. With these definitions in
tow, the mapping of bonds
\begin{eqnarray}
\label{bRR}
b_{(i,j)}&\dual& L_{(i,j)}, \quad \quad \quad \, 1 \le i,j \le N,  \\
R_{(i,j;1)}&\dual&S^\dagger_{(i,j-1;2)},~ 1 \le i \le N,~  2 \le
j \le N, \nonumber \\ 
R_{(i,j;2)}&\dual&S_{(i-1,j;1)},~ 2 \le i \le N, ~ 1 \le
j \le N, \nonumber
\end{eqnarray}
implements the duality transformation  \(H_{\sf ZG}\dual H_{\sf XY}\).
As  the operators \(R^{\:}_{(1,N+1;1)},R^{\dagger}_{(1,N+1;1)}\) {\it do not
appear} in \(H_{\sf ZG}\), the operator \(X_{(1,N+1;1)}\) constitutes a
{\it boundary symmetry} of \(H_{\sf ZG}\). Similar to the duality
between the  one-dimensional theories of Eqs. (\ref{HH2}) and
(\ref{hdh}), this is a {\it gauge-reducing} duality. The gauge
symmetries of \(H_{\sf ZG}\), given by \(A_{(i,j)}=R^{\:}_{(i,j;1)}
R^{\:}_{(i,j;2)}R^\dagger_{(i-1,j;1)}R^\dagger_{(i,j-1;2)}\), ~ \(2 \le i,j 
\le N\),  are removed by the mapping since \(A_{(i,j)}\dual \mathds{1}\). 

In the thermodynamic ($ N \to \infty$)  limit, the strongly-coupled ($J
\gg h$) phase of the XY model displays spontaneous symmetry breakdown of
its global \(U(1)\) symmetry with generator  \(L_{\sf
XY}=\sum_{i,j=1}^NL_{(i,j)}\), as evinced by a non-vanishing 
$\langle{\sf XY}|s_{\r}^{\;}s^\dagger_{\r'}|{\sf XY}\rangle$ in the limit
$ |\r - \r'| \rightarrow \infty$. By virtue of
being dual to the XY system, the gauge theory displays a non-analyticity
in its ground state energy as $h$ is varied and its symmetry is broken. 
However, the phase transition in the gauge
theory cannot be characterized by a local OP.  So, how can the duality
connecting the two models bridge the drastic gap separating the physical
interpretation of their common phase diagram?  The answer lies in our notion
of holography, since  
\begin{equation}
-X_{(1,N+1;2)}=\sum_{i,j=1}^Nb_{(i,j)}\dual L_{\sf XY}.
\end{equation}  
Thus, the global symmetry of the XY model is holographically  dual to
the (local) boundary symmetry $X_{(1,N+1;2)}$ of its dual gauge theory and  
cannot be spontaneously broken  in this dual theory \cite{elitzur}. 
This is how holographic symmetries explain the non-Landau nature of critical transitions 
in the \(\Z\) gauge theory. 
There are no edge modes nor localization associated with this holographic symmetry 
as the ordered phase of the XY model is gapless.

\begin{figure}[h]
\includegraphics[angle=0,width=.5\columnwidth]{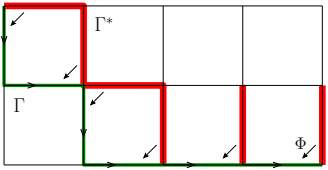}
\caption{Dual sets of links \(\Gamma^*\) and \(\Gamma\).}
\label{t_path_fig}
\end{figure}

We now derive a generalized OP for the \(\Z\) gauge theory. 
Let \(\Gamma\) be an {\it oriented} path from \(\r\) to \(\r'\) made of
directed links   \(\l\in \Gamma \), and we adopt the convention that \(S_\l\equiv S_{(\r;\mu)}\) if \(\l\) 
points from  \(\r\) to \(\r+\bm{e_\mu}\), or \(S_\l \equiv S_{(\r;\mu)}^\dagger\) if 
\(\l\) points oppositely from \(\r+\bm{e_\mu}\)  to \(\r\). Then
$s_{\r}^{\;}s_{\r'}^\dagger=\prod_{\l\in \Gamma} S_\l$.
Also let \(\Gamma^*\) denote the set of links \(\l^*\) 
such that \(\Phi_\d(R_{\l^*})=S_\l\) (\(\Gamma^*\) need not be 
continuous, see Fig. \ref{t_path_fig}). Then
\begin{equation}\label{string}
\langle {\sf ZG}| \prod_{\l^*\in\Gamma^*}R_{\l^*}|{\sf
ZG}\rangle\dual \langle {\sf XY}| s_{\r}^{\;}s_{\r'}^\dagger|{\sf XY}\rangle,
\end{equation} 
and so  {\it the 
string correlator on the left-hand side 
is a generalized OP for the \(\Z\) gauge theory, displaying long-range order in the ordered phase}.
On a {\it closed} path, \(\prod_{\l^*\in\Gamma^*}R_{\l^*}\) reduces to
a  product of gauge symmetries.

Finally, we couple the \(\Z\) gauge theory to a \(\Z\) matter field 
(defined on sites $\r$), 
$H_{\sf ZH}= H_{\sf ZG}+H_{\sf M}$, with 
\begin{equation}
H_{\sf M}=  \sum_\r\Big[\lambda(R_\r+R_\r^\dagger)+
\kappa\sum_{\mu=1,2} l_{(\r;\mu)}^2\Big],
\end{equation} 
and \(l_{(\r;\mu)}\equiv X_{\r+\bm{e_\mu}}-qX_{(\r;\mu)}-X_\r\).
The resulting matter-coupled theory $H_{\sf ZH}$  is dual to the 
Abelian Higgs model \cite{fradkin_shenker} with Hamiltonian
\begin{eqnarray}
&H_{\sf AH}&=\sum_\r\Big[\lambda(B_\r+B_\r^\dagger)+ h L_\r^2 \\
&&+\sum_{\mu=1,2}\Big(\kappa
L^2_{(\r;\mu)}+\frac{J}{2}(S^{(q)}_{(\r,\mu)}+S^{(q)\dagger}_{(\r,\mu)})\Big)\nonumber\Big].
\end{eqnarray}
Here \(S^{(q)}_{(\r,\mu)}\equiv
s^{\:}_\r s_{(\r;\mu)}^q s_{\r+\bm{e_\mu}}^\dagger\) includes a coupling
with integer charge \(q\) to the \(U(1)\) gauge field \(s_{(\r;\mu)}\equiv 
e^{-i\theta_{(\r;\mu)}}\), $s^q_{(\r;\mu)}\equiv 
e^{-iq\theta_{(\r;\mu)}}$, and \(B_\r\equiv 
s^{\:}_{(\r;1)}s^{\:}_{(\r+\i;2)}s_{(\r+\j;1)}^\dagger s_{(\r;2)}^\dagger\).
The correspondence between the two models, established by the mapping of bonds
\begin{equation}
\begin{array}{cc}
 \ \ \  R_\r\dual B_{\r-\i-\j}^\dagger,& 
\! \!\!\!\!\!\!\! \!\!b_\r\dual L_\r\\
\! \! \!  R_{(\r;1)}\dual S^{(q)\dagger}_{(\r-\j;2)}, 
& \! \! \!  R_{(\r;2)}\dual S^{(q)}_{(\r-\i,1)} \label{d2}\\
   l_{(\r;1)} \dual L_{(\r-\j;2)} ,&  \ \ \  l_{(\r;2)}\dual-L_{(\r-\i;1)} ,
\end{array}
\end{equation}
which holds only on physical gauge-invariant states. The reason is that
\(\Phi_\d\) preserves all commutation relations while
``trivializing'' all gauge symmetries. More precisely,  
\(H_{\sf ZH}\)'s gauge symmetries \(G_\r=R_\r A_\r\) map to
\(\Phi_\d(G_\r)=\mathds{1}\), while \(H_{\sf AH}\)'s gauge generators  
\(g_\r=L_{(\r;1)}+L_{(\r;2)}-L_{(\r-\i;1)}-L_{(\r-\j;2)}-qL_\r\) map to 
\(\Phi_\d^{-1}(g_\r)=0\) as follows from Eqs. \eqref{d2} 
(\(\Phi_\d^{-1}\) is the mapping obtained from Eqs. \eqref{d2} by reversing all the arrows).

If the \(\Z\) matter field is weakly coupled to the \(\Z\) gauge field,
the string correlator of Eq. \eqref{string} will still change analytic behavior
across transitions. Then from Eqs. \eqref{d2}
\begin{equation}
\label{zhah}
\langle {\sf ZH}|
\prod_{\l^*\in\Gamma^*}R_{\l^*}|{\sf ZH}\rangle
\dual \langle{\sf AH}|s_{\r}^{\;}s^\dagger_{\r'}
\prod_{\l \in \Gamma}s^q_\l|{\sf AH}\rangle,
\end{equation}
we obtain a witness correlator for the Abelian Higgs model 
that reduces
to a Wilson loop on closed contours (\(\r=\r'\))
(here \(s_\l^q=s_{(\r;\mu)}^q\) if a link \(\l\)
points from  \(\r\) to \(\r+\bm{e_\mu}\) and 
\(s_\l^q=s_{(\r;\mu)}^{q\dagger}\) otherwise).  
This non-local correlator is directly related to intuitively
motivated generalized OPs like 
\(\langle{\sf AH}|s_{\r}^{\;}s^\dagger_{\r'}
\prod_{\l \in \Gamma}s^q_\l|{\sf AH}\rangle/\langle{\sf AH}|
\prod_{\l \in \Gamma_C}s^q_\l|{\sf AH}\rangle\) conjectured in earlier work 
\cite{sondhi,bricmont,fm} (\(\Gamma_C\) denotes a closed loop roughly
twice as long as \(\Gamma\) and containing it).


{\it Outlook.}---
As demonstrated,   holographic symmetries and generalized OPs appear in
numerous systems once boundary conditions are properly accounted for in
the framework of  bond-algebraic dualities. By providing a systematic methodology 
and many examples,
our results might bring the theory of generalized OPs and topological
orders  to a new level of development closer to that of Landau's
theory.  More key problems need to be tackled. First, the sufficient
conditions  under which a given topological order may be mapped to a
Landau order  and {\it viceversa} should be understood. Second, the
problem of associating  effective field theories to generalized OPs
should be studied systematically.  

{\it Acknowledgements.} This work was supported by the Dutch 
Science Foundation NWO/FOM and an ERC Advanced Investigator grant,
and, in part, under grants No. NSF PHY11-25915 and CMMT 1106293.

\balancecolsandclearpage

\section{ Supplemental Material} 

{\it Introduction.}---
In the main text, we illustrated key ideas by examining several examples:
 (i) the quantum one-dimensional Ising matter coupled gauge theory (the \(\mathds{Z}_2\)  Higgs model)
and its dual quantum spin chain in an external field,
(ii) a two-dimensional quantum XY model and its solid-on-solid (\(\Z\)) type gauge theory dual, 
and (iii) the quantum
$U(1)$ matter coupled (or Abelian Higgs) gauge theory and its \(\Z\) matter field coupled gauge dual. 

In what follows, we illustrate how our concepts can be similarly worked out in other examples: 
(iv) the two-dimensional transverse field Ising
model and its quantum Ising gauge theory dual, (v) the self-dual two-dimensional quantum  \(\mathds{Z}_2\) Higgs model,
(vi) the two-dimensional quantum $p$-clock model and its \(\mathds{Z}_p\)  gauge theory dual, 
(vii) the two-dimensional \(\mathds{Z}_p\) Higgs  theory which we earlier illustrated to be 
self-dual for all $p$, (viii) the two-dimensional \(\mathds{Z}_p\)  Higgs theory and its Extended Toric Code dual, and (ix) the quantum XY model
on the kagome lattice and its \(\Z\) gauge theory dual. Taken together, these examples demonstrate how holographic symmetries and generalized
OPs and associated correlators can be systematically derived in disparate theories.

{\it The two-dimensional transverse field Ising model and
its quantum Ising gauge theory dual.}---
 As in the main text, we will perform dualities for finite size systems and explicitly label the Cartesian 
coordinates $1 \le i, j \le N$ of sites $\r = (i,j)$ on an $N \times N$ square lattice. The planar transverse-field Ising model 
for such any $N \times N$ square lattice system is given by
\begin{eqnarray}
\label{hi}
H_{\sf I}&=&-h\sum_{i,j=1}^N \sigma^x_{(i,j)}\label{ising_h}\\
&-&J\sum_{i=1}^{N-1}\sum_{j=1}^{N}\sigma^z_{(i,j)}\sigma^z_{(i+1,j)}-
J\sum_{i=1}^{N}\sum_{j=1}^{N-1}\sigma^z_{(i,j)}\sigma^z_{(i,j+1)}. \nonumber
\end{eqnarray}
The first term denotes the transverse field along the $x$ direction while last two terms correspond to Ising interactions (along the internal $\sigma^z$ direction) associated with horizontal and vertical links 
of the lattice respectively. This transverse field Ising model
is dual to (the gauge-invariant sector of) the \(\Z_2\) Ising gauge theory
in the infinite size limit \cite{wegnera}.
For open boundary conditions, we obtain via a bond-algebraic duality
the {\it exact} dual of \(H_{\sf I}\) which reads 
\begin{eqnarray}
\label{hig}
H_{\sf IG}&=&-h\sum_{i,j=1}^N B_{(i,j)} \\
&-&J\sum_{i=2}^N\sum_{j=1}^N\sigma^x_{(i,j;2)}
-J\sum_{i=1}^{N}\sum_{j=2}^N\sigma^x_{(i,j;1)}.\nonumber
\end{eqnarray}
Following our earlier conventions, the link indices \((i,j;\mu)=(\r;\mu),\ \mu=1,2\), that appear {\it explicitly}
in \(H_{\sf IG}\) labeling spins \(\sigma^x\) will be termed{\it allowed indices}, with all other
link indices being {\it forbidden}. This convention will be invoked to define the plaquette product $B_{(i,j)}$
which will omit forbidden links. The plaquette operators will be of a similar nature to those of Fig. \ref{dual_ising_fig} of the main text.
In the ``bulk'', the plaquette operator \(B_{(i,j)}\) reads 
\begin{equation}\label{plaquetteSM}
B_{(i,j)}\equiv \sigma^z_{(i,j;1)}\sigma^z_{(i+1,j;2)}\sigma^z_{(i,j+1;1)}\sigma^z_{(i,j;2)}.
\end{equation}
At the topmost left site (see Fig. \ref{dual_ising_fig}),
\(B_{(1,N)}=\sigma^z_{(1,N;1)}\sigma^z_{(2,N;2)}\sigma^z_{(1,N+1;1)}.\)
Hence \(B_{(1,N)}\) involves one spin \(\sigma^z_{(1,N+1;1)}\) labelled by a forbidden 
link index (see Fig. \ref{dual_ising_fig}). Elsewhere on the boundary of the square lattice, the remaining boundary plaquettes are determined by 
Eq. \eqref{plaquetteSM} {\it provided spins labelled by forbidden link indices are omitted}.

We now examine to the bond algebra- the set of algebraic relations between the interaction terms in the Hamiltonian of Eqs. (\ref{hi}, \ref{hig}).
This will allow us to establish a duality mapping between the transverse field Ising model and the Ising gauge theory. 
A natural partition of the transverse field Ising Hamiltonian $H_{\sf I}$ of Eq. (\ref{hi}) is into three type of bonds:
(a) the transverse fields $\{\sigma^{x}_{(i,j)}\}$, (b) horizontal Ising interactions $\{ \sigma^z_{(i,j)}\sigma^z_{(i+1,j)} \}$,
and (c) vertical Ising interactions $\{ \sigma^z_{(i,j)}\sigma^z_{(i,j+1)} \}$. The bond algebra of this system is exhausted by the following relations: \newline

(1) All bonds of type (b) and (c) commute amongst themselves and each other.  \newline

(2) Single site bonds of type (a) anticommute with the two-site bonds of type (b) [or (c)] whenever any two such bonds share a common site
on the lattice; this is a trivial consequence of the anticommutation relation $\ \{\sigma^{x}_{\r}, \sigma^{z}_{\r} \} =0$. \newline

(3) Whenever bonds of type (a) share no common lattice sites with bonds of type (b) or (c) then any such bonds commute
with one another. That this is so is readily seen as $[\sigma^{x}_{\r}, \sigma^{z}_{\r'}]=0$ whenever $\r \neq \r'$. \newline

(4) The square of any bond is the identity operator ($(\sigma^{x}_{\r})^{2} = (\sigma^{z}_{r'}) \sigma^{z}_{\r'+ \mu})^{2} =1$)). \newline

(5) The product of any horizontal/vertical bonds of types (b) and type (c) 
around any closed plaquette is equal to the identity operator.
That is, as $(\sigma^{\r})^{2} =1$ for any site $\r$, for all $2 \le i, j \le N$, 
\begin{eqnarray}
\mathds{1}= [\sigma^z_{(i,j)}\sigma^z_{(i-1,j)}][\sigma^z_{(i-1,j)}\sigma^z_{(i-1,j-1)}] \nonumber
\\ \times [\sigma^z_{(i-1,j-1)}\sigma^z_{(i,j-1)}] [\sigma^z_{(i,j-1)}\sigma^z_{(i,j)}]
\end{eqnarray}

Upon removing gauge redundancies,  an identical set of relations appears for the bonds of the Ising gauge theory of Eq. (\ref{hig}).
To underscore this, we partition Eq. (\ref{hig}) into three types of bonds as follows:
(a') the plaquette operators $\{B_{(i,j)}\}$,  (b') gauge fields $\sigma^{x}_{i,j;2}$ on vertical and (c') gauge fields $\sigma^{x}_{i,j;1}$ on longitudinal links. 
That similarity of algebraic relations underlies the duality mapping between the theories. That the two theories of Eqs. (\ref{hi}, \ref{hig}) are 
indeed dual to each other on an $N \times N$ square lattice (with of an size arbitrary $N$) 
is readily seen by the mapping of the bonds (a,b,c) $\to$ (a',b',c'), or
\begin{eqnarray}
\label{long_explanation_Ising}
B_{(i,j)}&\dual&\sigma^x_{(i,j)},\\
\sigma^x_{(i,j;1)}&\dual&\sigma^z_{(i,j-1)}\sigma^z_{(i,j)},\quad j=2,\cdots,N,\\
\sigma^x_{(i,j;2)}&\dual&\sigma^z_{(i-1,j)}\sigma^z_{(i,j)},\quad i=2,\cdots,N,
\end{eqnarray}
once the local gauge symmetries of the Ising gauge theories are
removed and set to unity,
\begin{eqnarray}
\label{long_removal}
A_{(i,j)}=\sigma^x_{(i,j;1)}
\sigma^x_{(i,j;2)}\sigma^x_{(i-1,j;1)}\sigma^x_{(i,j-1;2)} \dual \mathds{1}.
\end{eqnarray}
With the constraint of Eq. (\ref{long_removal}) implemented, redundant gauge degrees of freedom
are eliminated- applying a gauge transformation of the form of Eq. (\ref{long_removal}) does not
lead to a new state. 
As the mapping of Eqs. (\ref{long_explanation_Ising}, \ref{long_removal})
preserves all of the algebraic relations amongst the bonds in both theories
of Eqs. (\ref{hi}, \ref{hig}), this mapping implements an duality \(H_{\sf IG}\dual H_{\sf I}\)
for any $N \times N$ square lattice.
Similar to the main text, we briefly elaborate on 
the change of number of degrees of freedom once Eq. (\ref{long_removal}) is implemented
and how this leads to an identical number of degrees of freedom in both the transverse field Ising
and Ising gauge theory. The counting of the number of degrees of freedom is
identical to that of the $U(1)$ system examined in the main text. We repeat it anew
here for the Ising systems at hand. The Hamiltonian \(H_{\sf IG}\) describes \(2(N-1)N+1\) spins
and possesses \((N-1)^2\) gauge symmetries which are removed by the constraint of Eq. (\ref{long_removal}).
This leaves a total of \(2(N-1)N+1 - (N-1)^2=
N^2\) independent spins. This is precisely the number of spins in \(H_{\sf I}\).

Now, we turn to a quintessential feature seen by this duality- that of the holographic symmetry.
As, by virtue of our definitions above of forbidden indices, \(\sigma^z_{(1,N+1;1)}\) makes an appearance only in $B_{(1,N)}$ (see Fig. \ref{dual_ising_fig})
and $\sigma^{x}_{1,N+1;1)}$ appears
nowhere in the Hamiltonian of Eq. (\ref{hig}), the operator \(\sigma^z_{(1,N+1;1)}\) constitutes
a {\it boundary symmetry} of \(H_{\sf IG}\). What is this boundary symmetry's origin? 
The answer is afforded by a simple calculation, 
\begin{equation}
\label{long_holography_ising}
\sigma^z_{(1,N+1;2)}=\prod_{i,j=1}^NB_{(i,j)}\dual Q_{\sf I}.
\end{equation}  
{\it Thus, the boundary symmetry \(\sigma^z_{(1,N+1;2)}\) is holographic.}
As seen by Eq. (\ref{long_holography_ising}),  this symmetry is a ``hidden'' dual of the global \(\Z_2\) symmetry of the Ising model. 
Its recognition became possible with the specific choice of operators and use of bond algebras which suited to treat exactly for finite size systems
and their boundaries. 
 
We conclude our examination of this duality and its consequences by deriving the OP of the Ising gauge theory. 
As we will establish now, this OP will be given by string products- products of operators along connected links.
To write this OP, we need to set a few preliminaries.  Towards this end, as in the main text, we let \(\Gamma\) denote a path (set of concatenated links) in the Ising model's lattice, 
starting at site \(\r_1\) and ending at \(\r_2\) so that
\(
\prod_{(\r,\mu)\in \Gamma}\sigma^z_\r\sigma^z_{\r+\bm{e_\mu}}=
\sigma^z_{\r_1}\sigma^z_{\r_2}.
\)
We furthermore let \((\r^*,\bm{e_{\mu}}^*)\) denote that link in the lattice of the Ising gauge theory
such that \(\Phi_\d(\sigma^x_{(\r^*,\bm{e_{\mu}}*)})=
\sigma^z_\r\sigma^z_{\r+\bm{e_\mu}}\). 
From these definitions, it follows that
\begin{equation}
\label{IIG}
\mathcal{G}_{\Gamma^*}\equiv
\langle {\sf IG}|\prod_{\small (\r^*,\bm{e_{\mu}}*)\in \Gamma^*}\sigma^x_{(\r^*,\bm{e_{\mu}}*)}
|{\sf IG}\rangle \dual \langle {\sf I}|\sigma^z_{\r_1}\sigma^z_{\r_2}
|{\sf I}\rangle.
\end{equation}
In Eq. (\ref{IIG}), $| {\sf IG} \rangle$ and $| {\sf I} \rangle$ denote, respectively, the ground states of
the Ising gauge (Eq. (\ref{hig}))  and Transverse field Ising model (Eq. (\ref{hi})). 
Thus, putting all of the pieces together, the (infinite separation limit of the) string correlator  
\(\mathcal{G}_{\Gamma^*}\) is the generalized OP
of the Ising gauge theory. This OP is the dual of
the Landau order as it appears in Eq. (\ref{landau_op}). 
   
{\it The self-dual \(\Z_2\) Higgs model in two dimensions.}---
We next examine the self-dual quantum Ising matter coupled gauge theory on the square lattice.  
Similar to its one-dimensional variant of Eq. (\ref{HH2}), the Hamiltonian of
this system is given by
\begin{eqnarray}
\label{2DZ2Higgs}
H_{\sf H}&=&-\sum_\r (hB_\r+\lambda\sigma^x_\r)\\
&-&\sum_\r\sum_{\mu=1,2}(J\tau^x_{(\r;\mu)}+\kappa\sigma^z_\r\tau^z_{(\r;\mu)}.
\sigma^z_{\r+\bm{e_\mu}}).
\end{eqnarray}
Similar to the main text, In the $\Z_2$ Higgs Hamiltonian of Eq. (\ref{2DZ2Higgs}), 
$\sigma^{x,z}_{\r}$ are Pauli operators that reside on the lattice site $\r$ (which play the role
of quantum matter fields) while the quantum gauge fields 
$\tau^{x,z}_{(\r;\mu)}$ are Pauli operators that lie on a square lattice link that connects the site $\r$ with a neighboring site site $\r+ \mu$ along
one of the two ($\mu = 1,2)$ square lattice directions. Following all that we have stated thus far, the reader can verify that 
the correspondence of bonds
\begin{eqnarray}
\quad B_\r&\sdual&\sigma^x_{\r},\\
\sigma^x_{(\r;1)}&\sdual&\sigma^z_{\r-\j}\tau^z_{(\r-\j;2)}\sigma^z_{\r},\label{sd1}\\
\sigma^x_{(\r;2)}&\sdual& \sigma^z_{\r-\i}\tau^z_{(\r-\i;1)}\sigma^z_{\r},\label{sd2}
\end{eqnarray}
along with analogous relations for the matter fields \cite{cona} proves that \(H_{\sf Higgs}\)
is self-dual in its gauge-invariant sector. It is straightforward
to check that \(\Gamma^*\) maps to the path \(\Gamma\) just as before under the self-duality
mapping of Eqs. (\ref{sd1},\ref{sd2}). Hence,
\begin{equation}
\label{z2fm}
\langle H | \prod_{(r; \mu) \in \Gamma^*} \sigma^{x}_{\r; \mu} | H \rangle \sdual \langle {\sf H}| \sigma^z_{\r_1}
\sigma^z_{\r_2} \prod_{(\r;\mu)\in \Gamma} \sigma^z_{(\r;\mu)}|{\sf H}\rangle,
\end{equation}
where $|H \rangle$ denotes the ground state of the Higgs Hamiltonian of Eq. (\ref{2DZ2Higgs}).
The gauge-invariant correlator on the right-hand side of Eq. (\ref{z2fm}) 
is related to known string OPs for Higgs theories \cite{bricmonta,fma}.
The reader is referred to these works for details concerning asymptotic behaviors of
string operators in the confined and deconfined phases. For completeness, we briefly reiterate that 
in these OPs \cite{bricmonta,fma}, the expectation value of Eq. (\ref{z2fm}) is divided by string \cite{bricmonta} (or Wilson loop \cite{fma} products
which are derived from $\Gamma$. This division enables the extraction of the non-trivial asymptotic large $\Gamma$ behavior of the string correlator of Eq. (\ref{z2fm});
This factor is, equivalently, related to the normalization of the state generated, in an imaginary time representation, by the application of the string product on
the ground state \cite{fma,sondhia}. Other relations to effective line tension resulting from combined bare line (borne by matter coupling) and surface (generated by pure gauge fields) tensions are discussed 
in \cite{sondhia}. 

{\it The \(\Z_p\) Gauge Model.}---
We may place the \(\Z_2\) theories of the previous 
sections and the \(U(1)\) theories in the main part of the paper at the ends
of an infinite sequence of models with \(\Z_p\) symmetry, \(p=2,3,\cdots, \infty\). 

A ``clock model'' spin is associated to the group \(\Z_p\) is described by two unitary 
\(p\times p\) 
matrices \(U, V\) satisfying \(U^p=\mathds{1}=V^p\) and \(VU=\omega UV\), with 
\(\omega=e^{i2\pi/p}\) a \(p\)th root of unity \cite{p-clock}. When $p=2$, these correspond to the Pauli operators $\sigma^{x,z}$. 
Thus the Hamiltonian for the quantum
\(p\)-clock model can be written as 
\begin{eqnarray}
\label{pc}
H_{\sf pC}=-\Big[\frac{J}{2}\sum_{i=1}^{N-1}\sum_{j=1}^{N}U_{(i,j)}U^\dagger_{(i+1,j)}
\ \ \ \ \ \ \\
+\frac{J}{2}\sum_{i=1}^{N}\sum_{j=1}^{N-1}U_{(i,j)}U^\dagger_{(i,j+1)}
+\frac{h}{2}\sum_{i,j=1}^N V_{(i,j)}\Big]+h.c.\label{pclock_h}\ .\nonumber
\end{eqnarray}
Similar to the $\Z_2$ theories, the \(p\)-clock is dual to (the gauge-invariant sector of) the 
\(\Z_p\) gauge theory  
\begin{eqnarray}
\label{pg}
H_{\sf pG}=-\Big[\frac{J}{2}\sum_{i=2}^N\sum_{j=1}^N V_{(i,j;2)}\\
+\frac{J}{2}\sum_{i=1}^{N}\sum_{j=2}^NV_{(i,j;1)}
+\frac{h}{2}\sum_{i,j=1}^N B_{(i,j)}\Big]+h.c.\ .\nonumber
\end{eqnarray}
Following, verbatim, the same convention as earlier (as also marked in Fig. \ref{dual_ising_fig}))
yet now applied to these operators,
the plaquette operator \(B_{(i,j)}\) for plaquettes in the ``bulk'' is given by
\begin{equation}\label{plaquetteUV}
B_{(i,j)}\equiv U_{(i,j;1)}U_{(i+1,j;2)}U^\dagger_{(i,j+1;1)}U^\dagger_{(i,j;2)}.
\end{equation}
On the boundary, 
\(B_{(1,N)}=U_{(1,N;1)}U^\dagger_{(2,N;2)}U^\dagger_{(1,N+1;1)}.\)
The remaining boundary plaquettes are determined by 
Eq. \eqref{plaquetteUV} provided operators labelled by forbidden link indices 
(those extend beyond the system boundaries) are omitted, see Fig. \ref{dual_ising_fig})).
The correspondence of bonds
\begin{eqnarray}
B_{(i,j)}&\dual& V^{\dagger}_{(i,j)},\\
V_{(i,j;1)}&\dual& U^\dagger_{(i,j-1)} U_{(i,j)}  ,\quad j=2,\cdots,N,\\
V_{(i,j;2)}&\dual& U_{(i-1,j)} U^\dagger_{(i,j)}  ,\quad i=2,\cdots,N,
\end{eqnarray}
establishes the (gauge-reducing) duality \(H_{\sf pG}\dual H_{\sf pC}\).
The operator \(U^\dagger_{(1,N+1;1)}\) is 
an {\it holographic symmetry} of \(H_{\sf pG}\), since
\begin{equation}
U^\dagger_{(1,N+1;2)}=\prod_{i,j=1}^NB_{(i,j)}\dual \prod_{i,j=1}^N V_{(i,j)}^{\dagger}.
\label{holpc}
\end{equation} 
The operator on the right-hand side of Eq. \eqref{holpc} 
is the {\it global} \(\Z_p\) symmetry of the quantum \(p\)-clock
model of Eq. (\ref{pc}). As we see from the left-hand side of Eq. \eqref{holpc}, this operator is dual to the {\it boundary symmetry} of the gauge theory, $U^\dagger_{(1,N+1;2)}$. 
Thus, this symmetry, $U^\dagger_{(1,N+1;2)}$, is holographic. 

As we have in earlier examples, we now systematically devise an order parameter for
the gauge theory. This Abelian gauge theory exhibits, as most others, both confined and deconfined phases. The program for methodically
deriving a generalized OP diagnostic will replicate our calculations in earlier
examples. First, we write down a Landau correlator
which directly captures the long range order in the Landau type system (in this case,
the clock model of Eq. (\ref{pc})). Relying on the duality between the quantum clock model
and the quantum $\Z_{p}$ gauge theory, we then determine the corresponding 
correlator which serves as a generalized OP for the gauge theory of Eq. (\ref{pg}).

The Landau theory of the quantum p-clock exhibits long range order as seen by the finite value of the two-point correlation function,
in the limit of infinite separation, 
\begin{equation}
\langle{\sf pC}|U_{(i,j)}U^\dagger_{(i',j')}|{\sf pC}\rangle,\quad 
\sqrt{(i-i')^2+(j-j')^2}\rightarrow \infty,
\end{equation}
where $| {\sf pC} \rangle$ denotes the ground state of the p-clock model of Eq. (\ref{pc}). 
We now map this two-point correlator onto a generalized OP for the gauge theory by invoking the duality that we found. 
Towards this end, our steps will replicate, yet once again, those that we invoked in earlier examples. Let \(\Gamma\) be an {\it oriented}
path from \(\r\) to \(\r'\), and \(\l\in \Gamma \) its links oriented accordingly. Furthermore, we employ the convention that \(U_\l=U_\r U_{\r+\bm{e_\mu}}^\dagger\) if \(\l\) points from 
\(\r\) to \(\r+\bm{e_\mu}\),
and \(U_\l=U_\r^\dagger U_{\r+\bm{e_\mu}}\) if \(\l\) points oppositely from \(\r+\bm{e_\mu}\) 
to \(\r\). With this conventions, 
\begin{equation}\label{decomposition}
U_{\r}U_{\r'}^\dagger=\prod_{\l\in \Gamma} U_\l.
\end{equation}
If we define \(\Gamma^*\) as the set
of links mapping to the path \(\Gamma\) under duality and \(\Phi_\d(V_{\l^*})=U_\l\),
then we will find that 
\begin{equation}\label{pstring}
\mathcal{G}=\langle {\sf pG}|
\prod_{\l^*\in\Gamma^*}V_{\l^*}|{\sf pG}\rangle\dual \langle {\sf pC}|
U_{\r}U_{\r'}^\dagger|{\sf pC}\rangle,
\end{equation} 
where $| {\sf pG} \rangle$ is the ground state of the gauge theory $H_{\sf pG}$.
In the confined phase of the gauge theory, the generalized OP $\mathcal{G}$ assumes a 
non-zero value in the limit of infinite $\Gamma^*$. 
By contrast, in its deconfined phase, the gauge theory of Eq. (\ref{pg}) has a vanishing 
value of $\mathcal{G}$ as $\Gamma^*$ is made asymptotically 
large. This result constitutes a generalization of our earlier Ising (i.e., $p=2$) results.
Thus, similar to our earlier examples, putting all of the pieces together, the 
(infinite separation limit of the) string correlator  
\(\mathcal{G}\) is a generalized OP for the \(\Z_p\) gauge theory distinguishing
confinement from deconfinement, and derived
from simple considerations of spontaneous symmetry breakdown, duality, 
and holographic symmetries.

{\it The self-dual \(\Z_p\) Higgs model.}---
We now extend the earlier Ising results to the $\Z_p$ system, review
the self-duality of the $\Z_p$ gauge theory
when it is coupled to $\Z_p$ matter fields, and derive new string operators for this
theory. This matter coupled
gauge theory is the quantum $\Z_p$ Higgs theory. 
That is, we will augment the (pure) gauge theory \(H_{\sf pG}\)  of Eq. (\ref{pg})
by terms that minimally couple the gauge fields on the lattice links $U_{\r, \mu}$ to matter
fields $U_{\r}$ and $U_{\r + \bm{e_\mu}}$ at sites that form the endpoints of those links
as well as by additional matter only terms $V_{\r}$,
\begin{equation}
\label{matterc}
H_{\sf matter} = -\sum_\r\Big[\frac{1}{2}V_\r+
\frac{\kappa}{2}\sum_{\mu=1,2} U_\r U^q_{(\r,\mu)}U_{\r+\bm{e_\mu}}^\dagger\Big]+h.c..
\end{equation} 
From the minimal coupling term, where $q$ is the power of the gauge field on the link that is 
coupled to the matter field,  similar to the $U(1)$ example in the main text, it is evident that $q$ 
plays the role of a charge. As \(U^{q}=U^{(p-q)\dagger}\), this charge can assume the values \(q=1,\cdots, Int[p/2]\) (where \(Int[\cdot]\) denotes the integer
par of its argument). On augmenting the Hamiltonian of Eq. (\ref{pg}) by the terms in Eq. (\ref{matterc}), the resulting model Hamiltonian \(H_{\sf pH}\)- 
the (charge \(q\)) \(\Z_p\) Higgs model- is self-dual
in its gauge invariant sector \cite{cona}. Specifically, this self-duality is given by the mapping 
\begin{eqnarray}
B_{\r}&\sdual& V_\r^\dagger\\
V_{(\r,1)}&\sdual& U_{\r-\j}^\dagger U_{(\r-\j,2)}^{q\dagger}U_\r,\label{psd1}\\
V_{(\r,2)}&\sdual& U_{\r-\i}U_{(\r-\i,1)}^qU_\r^\dagger, \label{psd2}
\end{eqnarray}
for the gauge fields, and the transformation
\begin{eqnarray}
V_\r^\dagger&\sdual& B_{\r-\i-\j}^\dagger\\
U_\r U_{(\r,1)}^{q}U_{\r+\i}^\dagger&\sdual& V_{(\r-\j,2)}^\dagger,\\
U_\r^\dagger U_{(\r,2)}^{q\dagger}U_{\r+\j}&\sdual& V_{(\r-\i,1)}^\dagger,
\end{eqnarray}
for the matter fields. 

The string operator in Eq. \eqref{pstring}, becomes, on invoking Eqs. (\ref{psd1}, \ref{psd2}), 
\begin{equation}\label{pfm}
\langle {\sf pH}|
\prod_{\l^*\in\Gamma^*}V_{\l^*}|{\sf pH}\rangle
\sdual \langle{\sf pH}|U_{\r}U^\dagger_{\r'}
\prod_{\l \in \Gamma}U^q_\l|{\sf pH}\rangle.
\end{equation}
In Eq. (\ref{pfm}), $| {\sf pH} \rangle$ denotes the ground state of the Higgs Hamiltonian (given by $(H_{\sf pG} + H_{\sf matter}$)- the sum 
of the Hamiltonians in Eqs. (\ref{pg}, \ref{matterc})). Similar to our earlier conventions, in Eq. (\ref{pfm}) we set \(U_\l=U_{(\r,\mu)}^q\) if \(\l\) points from \(\r\) to \(\r+\bm{e_\mu}\),
and \(U_\l=U_{(\r,\mu)}^{q\dagger}\) otherwise.
 
As in our discussions for the other examples of Higgs theories concerning string OPs and their normalization, in the infinite separation (i.e., $\Gamma$) limit, 
the gauge-invariant correlator on the right-hand side is related to a normalized
generalized OP proposed in Ref. \cite{bricmonta,fma}.
Taken on a {\it closed} path, the right-hand side of Eq. \eqref{pfm}
reduces to a Wilson loop.

{\it The \(\Z_p\) Extended Toric Code model.}---
The Toric code model \cite{kitaev1a} and its extensions, e.g., \cite{cona,vidala,tka} have gained much interest in recent years.
Their original motivation was to serve as a caricature of toy models in which some of the basic concepts of
topological quantum computing can be explained \cite{kitaev1a}. Since the model and its variants have been investigated heavily
large on their own right. The original Toric code model had an Ising symmetry. In the current section, we will further employ its $\Z_p$ variant.
 
In the current context, our interest in this system derives from the fact that the \(\Z_p\) Higgs model of the previous section is dual
to a generalization of the extended \(\Z_p\) Toric Code model (pETC).  This model is given by
\begin{eqnarray}
\label{petca}
H_{\sf pETC}= -\sum_\r\Big[\frac{1}{2}A_\r^{1/q}+\frac{h}{2}B_\r\\
+\sum_{\mu=1,2} (\frac{J}{2}V_{(\r,\mu)}+ \frac{\kappa}{2} U^q_{(\r,\mu)})
\Big]+h.c.\ ,
\end{eqnarray}
on the square lattice,
with \(A_\r=V_{(\r,1)}V_{(\r,2)}V_{(\r-\i,1)}^\dagger V_{(\r-\j,2)}^\dagger\). 
The standard extended Toric Code is recovered for \(p=2\), in which case
\(q=1\) is the only possible value. The \(\Z_p\) toric code discussed
in Ref. \cite{vidala} is recovered for \(q=1\). 

A gauge-reducing 
duality connecting \(H_{\sf pH}\) to \(H_{\sf pETC}\)  is given by 
\begin{eqnarray}
\label{map_petc}
V_\r \dual A_\r^{1/q},  \nonumber \\
U_\r U_{(\r,1)}^{q}U_{\r+\i}^\dagger&\dual& U_{(\r,1)}^{q},\\
U_\r^\dagger U_{(\r,2)}^{q\dagger}U_{\r+\j}&\dual& U_{(\r,2)}^{q\dagger}, \nonumber
\end{eqnarray}
while leaving the gauge fields invariant , \(B_\r\dual B_\r\) and 
\(V_{(\r,\mu)}\dual V_{(\r,\mu)}\). This duality removes degrees of freedom that lie
at lattice sites. While the Hamiltonian \(H_{\sf pETC}\) has no global
symmetries, the \(\Z_p\) Higgs model has one global symmetry. 
Similar to our earlier discussions, the duality \(H_{\sf pH}\dual  H_{\sf pETC}\) maps
a global symmetry of \(H_{\sf pH}\) to a holographic symmetry of 
\(H_{\sf pETC}\). 

The generalized OPs introduced in the previous section for the 
\(\Z_p\) Higgs model can be mapped to (dual) generalized OPs for the (self-dual) extended
Toric Code model. Employing Eqs. (\ref{map_petc}) on the correlator of Eq. (\ref{pfm}), 
\begin{equation}
\label{ph+}
\langle {\sf pH}|
\prod_{\l^*\in\Gamma^*}V_{\l^*}|{\sf pH}\rangle
\dual\langle {\sf pETC}|
\prod_{\l^*\in\Gamma^*}V_{\l^*}|{\sf pETC}\rangle,
\end{equation}
and
\begin{equation}
\label{ph-}
\langle{\sf pH}|U_{\r}U^\dagger_{\r'}
\prod_{\l \in \Gamma}U^q_\l|{\sf pH}\rangle
\dual \langle{\sf pETC}| \prod_{\l \in \Gamma}U^q_\l|{\sf pETC}\rangle,
\end{equation}
where $|{\sf pETC} \rangle$ is the ground state of the extended toric code model of Eq. (\ref{petca})
and, as in Eq. (\ref{pfm}), $ |{\sf pH} \rangle$ denotes the ground state of the Higgs model.
As \(H_{\sf pH}\) is self-dual in its gauge-invariant sector, \(H_{\sf pETC}\)
is indeed also self-dual. The two string operators of Eqs. (\ref{ph+},\ref{ph-})
transform into each other under this self-duality transformation. 

\begin{figure}[h]
\includegraphics[angle=0,width=.7\columnwidth]{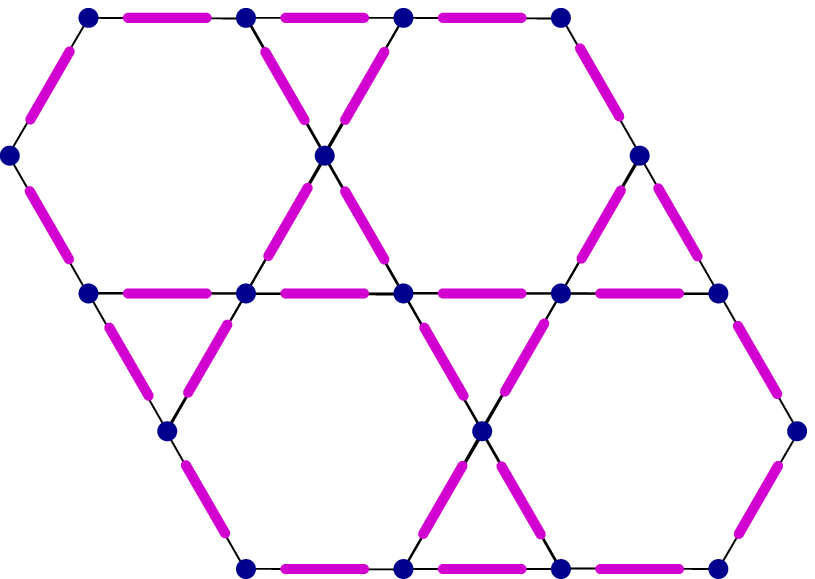}\\
\vspace{.2cm}
\includegraphics[angle=0,width=.8\columnwidth]{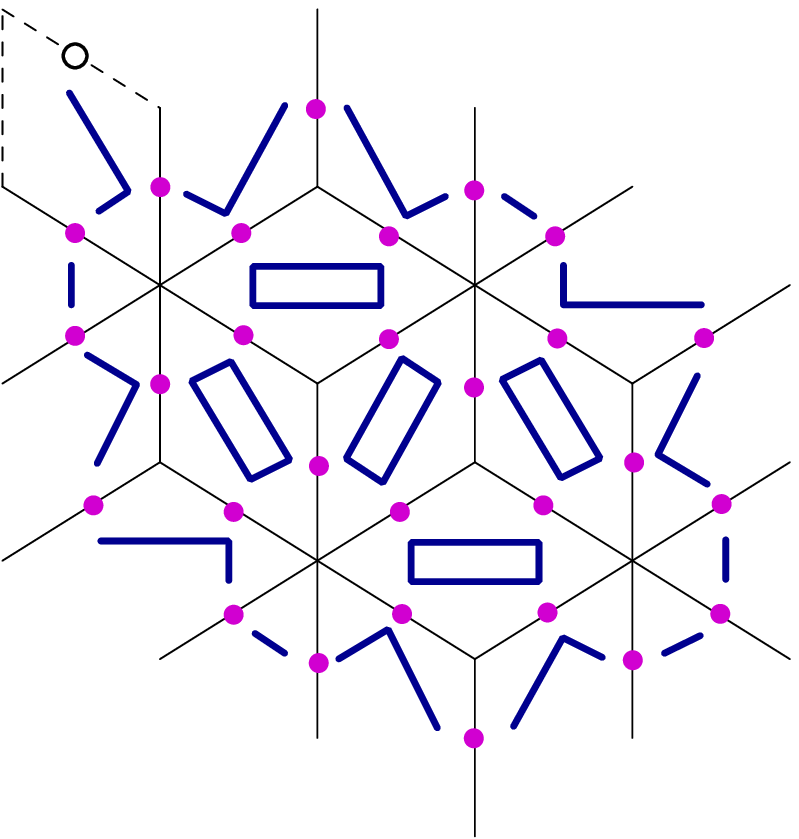}
\caption{The quantum, \(U(1)\) symmetric XY model on the kagome lattice 
is holographically dual to a solid-on-solid-like gauge theory on the dice lattice.}
\label{kagomes_fig}
\end{figure}

{\it  The quantum XY model on the kagome lattice.}---
We now examine a frustrated spin system with a continuous symmetry:
the quantum XY model on the open kagome lattice [see  Fig. \ref{kagomes_fig},we will denote this lattice by \(\Lambda_K\)],
\begin{equation}
H_{\sf XYK}=\sum_{\r\in\Lambda_K}\frac{ 1}{2I}L_\r^2+\sum_{\langle \r, \r'\rangle}
\kappa \cos(\theta_\r-\theta_{\r'}).
\end{equation}
The XY rotor at any kagome lattice site $\r$ has an orientation $\theta_{\r}$ and
\(L_\r=i\frac{\partial}{\partial\theta_\r}\). This XY Hamiltonian is invariant under
global rotations generated by the total angular momentum \(L=\sum_{\r\in\Lambda_K}
L_\r\). A duality maps this model to a solid-on-solid like ($\Z$) gauge theory 
(similar to the one discussed in the main text) which now appears on the
dice lattice \(\Lambda_{ D}\),
\begin{equation}
H_{\sf XYK}^D=\sum_{\l \in \Lambda_{ D}}\lambda (R_\l+R_\l^\dagger)+
\sum_{\p \in \Lambda_{ D}}\frac{ 1}{2I}\Big(\sum_{\l\in\partial \p} (-1)^{\l}
X_\l\Big)^2.
\end{equation} 
Here, \(\l\) are the oriented links of the dice lattice, \(\p\) its plaquettes,
and \((-1)^{\l}=\pm 1\), with  \((-1)^{\l}=-1\) only if \(\l\) viewed as a link in 
\(\Lambda_{ D}\) has opposite orientation to that assigned by  \(\l\in\partial \p\) 
(\(\partial \p=\) oriented boundary of \(\p\)). Similar to the examples that we discussed earlier, the
exact dual possesses incomplete plaquette at the boundaries; this is seen in the lower panel
of the figure. Under the duality transformation, the generator of the global rotational symmetry of the 
XY model maps onto a holographic symmetry,  
\begin{equation}
L=\sum_{\r\in\Lambda_K}L_r\dual \sum_{\p\in\Lambda_D} \sum_{\l\in\partial \p} (-1)^{\l}
X_\l = X_{\l'}.
\end{equation} 
Here, \(X_{\l'}\) is a degree of freedom placed at the boundary link marked with an open
circle. Any other \(X_{\l}\) enters the twice sum, with opposite sign. 
This is then an example of holography for a continuous symmetry in a highly frustrated system. 

\begin{thebibliography}{}

\bibitem{book}
H. Nishimori and G. Ortiz, {\it Elements of Phase Transitions and
Critical Phenomena}  (Oxford University Press, Oxford, 2011). 

\bibitem{batista} 
C. D. Batista, G. Ortiz, and J. E. Gubernatis, Phys. Rev. B {\bf 65},
180402(R) (2002).

\bibitem{furukawa}
S. Furukawa, G. Misguich, and M. Oshikawa, Phys. Rev. Lett. {\bf 96},
047211 (2006).

\bibitem{wen}
X-G. Wen, Int. J. Mod. Phys. B {\bf 4}, 239 (1990).

\bibitem{TQO} 
Z. Nussinov and G. Ortiz, Annals of Physics {\bf 324}, 977 (2009); 
Proceedings of the National Academy of Sciences {\bf 106},  16944 
(2009).

\bibitem{sondhi}
K. Gregor, D. A. Huse, R. Moessner, and S. L. Sondhi, New J.Phys. {\bf
13}, 025009 (2011).
         
\bibitem{kitaev1}  
A. Yu Kitaev, Ann. Phys. {\bf 303}, 2 (2003). 

\bibitem{kitaev2}  
A. Yu Kitaev, Ann. Phys. {\bf 321}, 2 (2006).

\bibitem{wegner}
F. J. Wegner, J. Math. Phys. {\bf 12}, 2259 (1971). 

\bibitem{con}
E. Cobanera, G. Ortiz, and Z. Nussinov, Phys. Rev. Lett. {\bf 104}, 020402 (2010);
Adv. Phys. {\bf 60}, 679 (2011).

\bibitem{nadprl}
E. Cobanera, G. Ortiz, and E. Knill, arXiv:1206.1367v1 [cond-mat.stat-mech]
(2012).

\bibitem{bond} 
Z. Nussinov and G. Ortiz, Phys. Rev. B {\bf 79},  214440 (2009).

\bibitem{zon}
Z. Nussinov, G. Ortiz, and E. Cobanera, Ann. Phys. {\bf 327}, 2491 (2012).

\bibitem{bricmont}
J. Bricmont and J. Fr\"olich, Phys. Lett. B {\bf 122}, 73 (1983).

\bibitem{fm}
K. Fredenhagen and M. Marcu, Phys. Rev. Lett. {\bf 56}, 223 (1986);
Comm. Math. Phys. {\bf 92}, 81 (1983).

\bibitem{thermal} 
Z. Nussinov and G. Ortiz, Phys. Rev. B {\bf 77},  064302 (2008).




\bibitem{kitaev_wire}
A. Yu Kitaev, Phys.-Usp. {\bf 44}, 131 (2001).

\bibitem{SI}
Supplemental material. 



\bibitem{fradkin_shenker} 
E. Fradkin and S. H. Shenker, Phys. Rev. D {\bf 19}, 3682 (1979).

\bibitem{vidal}
M. D. Schulz, S. Dusuel, R. Orus, J. Vidal, and K. P. Schmidt, 	
New J. Phys. 14, 025005 (2012).

\bibitem{elitzur}
S. Elitzur, Phys. Rev. D {\bf 12}, 3978 (1975).

\bibitem{herbut}
I. Herbut, {\it A Modern Introduction to Critical Phenomena} (Cambridge University Press,
Cambridge, 2010).

\bibitem{commutators}
A precise estimate follows from 
\([H_{\sf GK}[\kappa=0],\Gamma_1]=-2iJ(-h/J)^Nb_N\sigma^z_{(N-1;1)}\cdots\sigma^z_{(1;1)}\),
and \([H_{\sf GK}[\kappa=0],\Gamma_2]=-2iJ(h/J)^Na_1\sigma^z_{(1;1)}\cdots\sigma^z_{(N-1;1)}\).

	

\end{thebibliography}

\begin{thebibliography}{}

\bibitem{wegnera}
F. J. Wegner, J. Math. Phys. {\bf 12}, 2259 (1971). 

\bibitem{bricmonta}
J. Bricmont and J. Fr\"olich, Phys. Lett. B {\bf 122}, 73 (1983).

\bibitem{fma}
K. Fredenhagen and M. Marcu, Phys. Rev. Lett. {\bf 56}, 223 (1986);
Comm. Math. Phys. {\bf 92}, 81 (1983).



\bibitem{sondhia}
K. Gregor, D. A. Huse, R. Moessner, and S. L. Sondhi,
New J.Phys. {\bf 13}, 025009 (2011).


\bibitem{p-clock}
G. Ortiz, E. Cobanera, an Z. Nussinov, 
     Nuclear Physics B {\bf 854}, 780 (2011). 
     
     \bibitem{cona}
E. Cobanera, G. Ortiz, and Z. Nussinov, Phys. Rev. Lett. {\bf 104}, 020402 (2010);
Adv. Phys. {\bf 60}, 679 (2011).

\bibitem{kitaev1a}  A. Kitaev, Ann. Phys. {\bf 303}, 2 (2003). 

\bibitem{vidala}
M. D. Schulz, S. Dusuel, R. Orus, J. Vidal, and K. P. Schmidt, 	
New J. Phys. 14, 025005 (2012).

\bibitem{tka}
I.S. Tupitsyn, A. Kitaev, N.V. Prokof'ev, and P.C.E. Stamp, 
Phys. Rev. B {\bf 82}, 085114 (2010).

\end{thebibliography}
\end{document}